# Oscillations and huge preferences of PbTe crystal surface sputtering under Secondary Neutral Mass Spectrometry conditions

D.M. Zayachuk[a], E.I. Slynko[b], V.E. Slynko[b], A. Csik[c]

[a] Lviv Polytechnic National University, Lviv, Ukraine
[b] Institute of Material Science NASU, Chernivtsy, Ukraine
[c] Institute for Nuclear Research, Hungarian Academy of Sciences, Debrecen, Hungary

a b s t r a c t

Sputtering of PbTe crystals by $Ar^+$ plasma with ion energies 50–550 eV is investigated. A dependence of sputter yields of the Te and Pb on the sputtering ion energy and sputtering time is measured. New phenomena: aperiodical oscillations of Pb and Te sputtering; a huge preference of Te sputtering reaching more than two orders of magnitude at the beginning of sputtering process; and a significant excess of Te integrated sputter yield over that of Pb for prolonged sputtering by low energy plasma at 50–160 eV, are observed. It is substantiated that these phenomena can be caused by the peculiarities of the charge states of the interstitial Pb and Te in PbTe crystal matrix and the processes of re-deposition of sputtered atoms on the sputtered surface.

## 1. Introduction

Application of sputtering in Secondary Neutral Mass Spectrometry (SNMS) is known to be an effective method for composition analysis and depth profiling of multicomponent and doped solids [1–3]. Recently we successfully used this method for depth profiling across of Eu-doped PbTe single crystals grown from melt [4]. In SNMS method the composition of multicomponent systems is determined by measuring the relative content of the sputtered species. In this case, the problem of preferential sputtering arises and should always be considered [5]. According to linear cascade theory for $E»U$ ($E$ is the beam energy and $U$ is the surface binding energy) the ratio of partial sputter yields Y of 1 and 2 elements are:

$$\frac{Y_1}{Y_2} \simeq \frac{N_1^S}{N_2^S}\left(\frac{M_2}{M_1}\right)^{2m}\left(\frac{U_2}{U_1}\right)^{1-2m}. \quad (1)$$

Here $N_i^S$ and $M_i$ are the respective surface concentration and mass of the $i^{th}$ species, $m$ is the exponent in the low energy power cross section, a quantity close to zero [1]. Under prolonged sputtering the ratio of sputter yields monotonically tends to the ratio of the bulk element concentrations [5].

Herein, we focus on the specifics of the Pb and Te sputtering from PbTe crystal surface by Ar plasma under SNMS conditions.

## 2. Experiment

The crystals for investigations were grown from melt by the Bridgman method. The high purity Pb and Te, taken in stoichiometric ratio, were used for crystal growth. The lateral cylindrical surfaces of crystal were used for sputtering and analysis by SNMS method. Conditions for SNMS were the same as described in [4]. The analysis of the samples was performed in the direct bombardment mode by using $Ar^+$ ions with energy 50-550 eV. Ions were extracted from the Ar plasma by a negatively biased sample. The erosion area was confined to a circle of 2 mm in diameter by a Ta mask.

## 3. Results

Features of formation of the PbTe sputtered phase under impact of $Ar^+$ plasma in the SNMS conditions deserving special attention are the following.

- The intensity of Pb and Te signals varies during the prolonged sputtering, exhibiting the pronounced features of aperiodical oscillation process. Fig. 1 shows an example for the Te sputtering by $Ar^+$ plasma using different energies throughout one minute sputtering time.

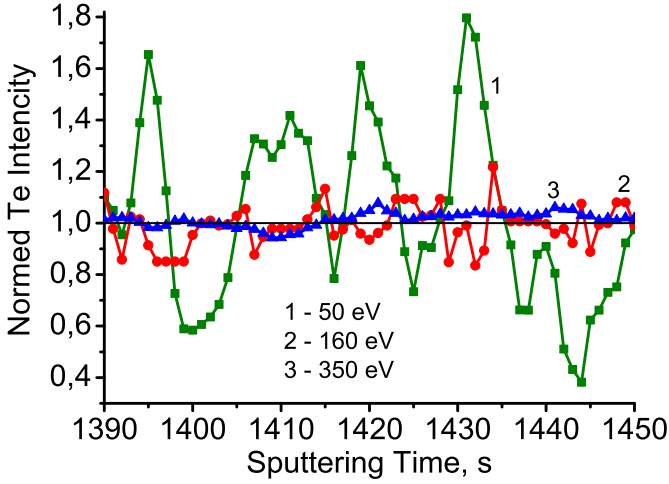

**Fig. 1.** Normalized by the mean Te signal vs. sputtering time, measured on PbTe lateral surface for different sputtering energies throughout one minute sputtering time.

- The amplitude of oscillations of the Pb and Te intensities increases when sputtering energy decreases. The maximum deviations of the measured magnitudes from their average values reached 80% for sputtering energy 50 eV and 10% for energy 550 eV.
- The average intensity of Pb and Te signals also varies during prolonged sputtering. It decreases with sputtering time at a sputtering energy of about 160 eV and less and increases at about 350 eV and higher (Fig. 2).
- A preferential sputtering process takes place during the sputtering, which depends on the energy of $Ar^+$ beam. At low sputtering energies of about 160 eV and less in the beginning of the process Te dominates in sputtered phase, but Pb dominates at high energies of about 350 eV and higher (Fig. 3 and Fig. 4).
- The degree of preferential Te sputtering at the lowest energy 50 eV is very large, exceeding two orders of magnitude (curve 1, Fig. 4). When Pb dominates in the sputtered phase, we observe much smaller differences between Pb and Te integrated sputter yields $Y_{tot}^{Pb}$ and $Y_{tot}^{Te}$ – they don't exceed approximately 12% (curves 3,4, Fig. 4).
- Ratio $Y_{tot}^{Te}/Y_{tot}^{Pb}$ oscillates during prolonged sputtering. Only for high sputtering energies 350 and 550 eV it tends to unity. For sputtering energies of about 160 eV and less ratio $Y_{tot}^{Te}/Y_{tot}^{Pb}$ does not approach unity even after almost one hour of sputtering (Fig. 4).

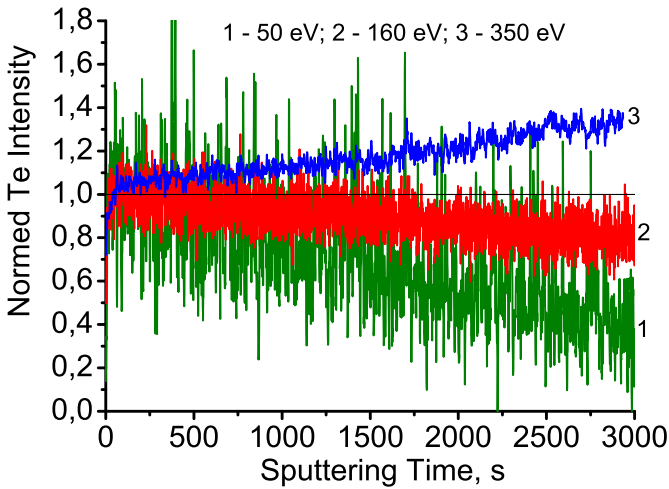

**Fig. 2.** Te signal intensity (normalized by the mean in the range from 10 to 100 sec) vs. sputtering time for different sputtering energies.

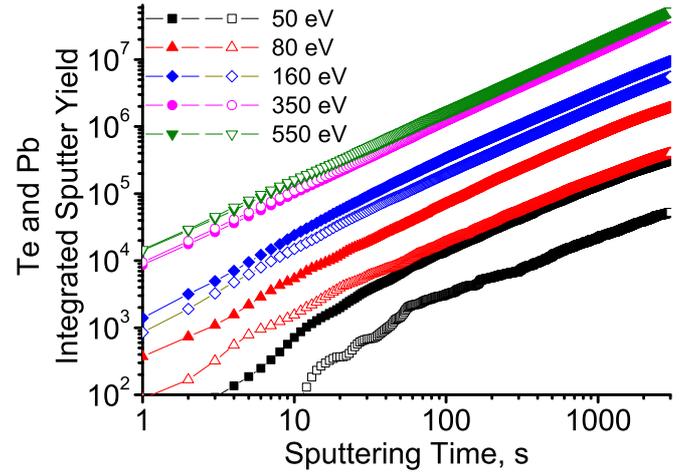

**Fig. 3.** Integrated sputter yields $Y_{tot}^{Pb}$ (open symbols), and $Y_{tot}^{Te}$ (solid symbols) vs. sputtering time for different sputtering energies.

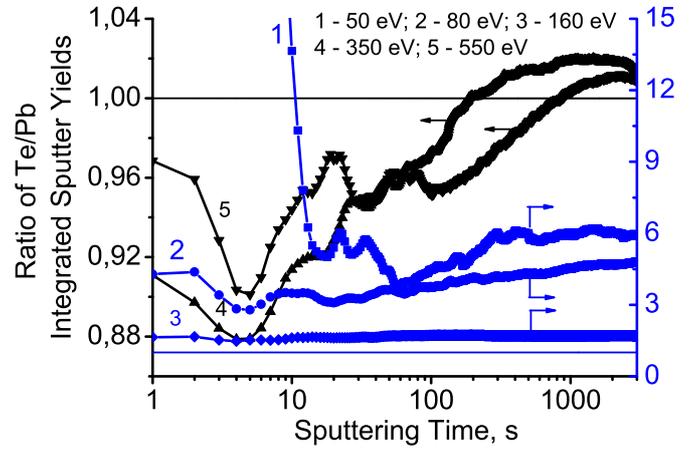

**Fig. 4.** Ratio of integrated sputter yields $Y_{tot}^{Te}/Y_{tot}^{Pb}$ vs. sputtering time for different sputtering energies.

## 4. Discussion

Let us first consider why the preference of Pb and Te sputtering changes over when sputtering energy changes. Ratio of Pb and Te partial sputter yields can be estimated using the Eq. (1).Their initial surface concentrations in PbTe crystals are equal. So, heavy Pb can dominate over light Te in the initial PbTe sputtered phase, exceeding the sputtering rate of Te at given sputtering energy, only if its surface binding energy is lower. That is what we are observing for sputtering with energy 350 and 550 eV (curves 4, 5, Fig. 4). Thus we conclude that surface binding energy of Pb in PbTe is indeed lower than that of Te.

Then why is the amount of Te higher in PbTe sputtered phase at low sputtering energies? We suggest that the reason lies in peculiarities of charge states of the interstitial defects in PbTe lattice. There interstitial Pb is double-charged donor with zero activation energy, and interstitial Te is electrically neutral [6]. Low energy sputtering around and below 160 eV is near-threshold sputter process. In this case, the primary recoil atoms dominate in sputtered flux [7,8]. Back-scattered neutral Te atom knocked out from an own site into an interstitial position is free to leave the PbTe sputtered surface. On the other hand, back-scattered $Pb^{2+}$ ion ejected from site into interstitial experiences difficulties in sputtering because it must overcome an additional potential barrier created by the positive potential of Ar plasma. Therefore, at low sputtering energies Te should significantly dominate in sputtered phase, despite the fact that its surface binding energy is higher than that of Pb. That is what we observe in the experiment (curves 1–3, Fig. 4).

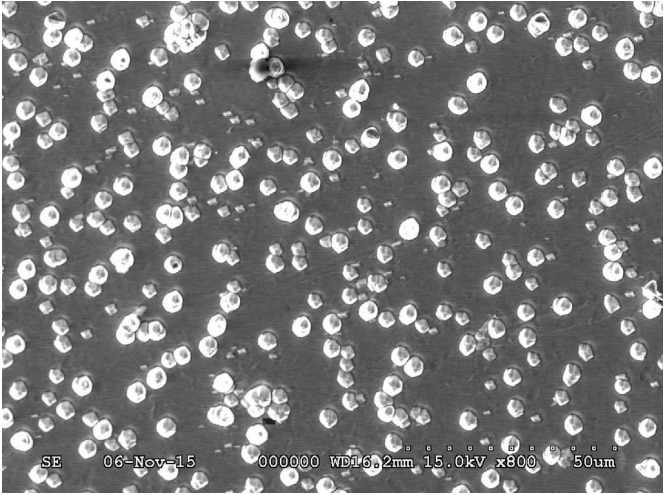

**Fig. 5.** Scanning electron microscope picture of PbTe crystal surface after sputtering by argon plasma with 160 eV ions energy for 3200 s.

Why do the Pb and Te sputter yields oscillate? To explain this unusual phenomenon we will consider the following. First, during prolonged sputtering the combined integrated sputter yield of material as a whole oscillates as well as Pb and Te sputter yields. Second, at any sputtering stage the surface of sputtered PbTe crystal is covered by the array of crystalline formations that are deposited on the sputtered surface during its sputtering (Fig. 5).

Study of the PbTe sputtered surface morphology is a subject of separate investigation, and its results will be published. Here we only emphasize two important facts. First, the density of crystalline formations on the sputtered surface at comparable sputter times increases when sputtering energy decreases. Second, the results of energy dispersive X-ray spectroscopy (EDX) measurements show that the composition of both newly formed crystallites and sputtered crystal surface is the same, to a first approximation. So, one can definitely assume that the sputtered phase of PbTe crystals is formed by the superposition of two simultaneous processes – material sputtering by the Ar plasma and deposition of the sputtered material on the sputtering surface. The first process increases the amount of sputtered material, the second decreases it.

As the sputtering process is preferential the sputtering rate of Pb and Te from the PbTe crystal surface will change over time. The sputtering of crystal surface leads to its enrichment by the component of PbTe crystal with lower sputter yield. This enrichment, in turn, leads to the increase of the other component sputter yield and the reduction of the first one from the crystal surfaces. Sputtering of surface of each newly deposited crystallite repeats the process.

Nucleation of the new formations on sputtering surface is a random process. Hence, the deposition rate of the sputtered material on the sputtering surface will not be constant too. Overlaying the two processes occurring at variable rates, one of which increases the amount of material in the sputtered phase while the other decreases it, can cause non-periodic oscillations of sputter yields of Pb and Te during prolonged sputtering of the PbTe surface, and that is what we are observing. As mentioned before the density of crystalline formations increases when sputtering energy decreases. So, nucleation and growth of the new formations on the sputtering surface is apparently responsible for decrease of Pb and Te sputter yields with time, as well as for significant deviation of $Y_{tot}^{Te}/Y_{tot}^{Pb}$ ratio from unity for large sputtering times at low energies of about 160 eV and less.

## 5. Conclusions

Summarizing, it should be emphasized that the presence of oscillatory sputtering of PbTe components will affect the determination of concentrations of various elements in the crystal matrix. To reduce this effect measurement is best done with high-energy plasma of about 300 eV or higher. On the other hand, sputtering at low energy of about 100 eV and lower under favorable conditions may be used for testing the difference of charge states of native interstitial atoms in crystal matrix.